\documentclass[3p,times,twocolumn]{elsarticle}
\usepackage{hyperref}
\usepackage{ecrc}

\volume{00}

\firstpage{1}

\journalname{Nuclear Physics B Proceedings Supplement}

\runauth{}

\jid{nuphbp}

\jnltitlelogo{Nuclear Physics B Proceedings Supplement}

\usepackage{amsmath,amssymb}
\usepackage{graphicx}
\usepackage{epsfig}
\usepackage{euscript}
\usepackage{pslatex}
\usepackage{xcolor}
\usepackage{fancybox}
\usepackage{enumerate}
\usepackage{widetext}

\newcommand{\Meu}{\EuScript{M}}
\newcommand{\Mmf}{\mathfrak{M}}

\newcommand{\KK}{${\cal KK}$}

\def\Order#1{${\cal O}(#1)$}
\def\OrderLL#1{${\cal O}(#1)_{\rm LL}$}

\newcommand{\sfac}{\mathfrak{s}}

\def\st{\hbox{}} 

\usepackage{amsthm}

\usepackage[figuresright]{rotating}

\begin{document}

\begin{frontmatter}



\dochead{\small\bf BU-HEPP-14-07, Aug., 2014}

\title{CEEX EW Corrections for $f \bar{f}\rightarrow f'\bar{f'}$ at LHC, Muon Colliders and FCC-ee as Realized in KK MC 4.22}


\author[label1]{B.F.L. Ward}
\author[label2]{S. Jadach}
\author[label2]{Z. Was}

\address[label1]{Baylor University, Waco, TX, USA}
\address[label2]{Institute of Nuclear Physics, Cracow, PL}

\begin{abstract}
With an eye toward the precision physics of the LHC, FCC-ee and possible high energy muon colliders, we present the extension of the CEEX (coherent exclusive
exponentiation) realization of the YFS approach to resummation in our KK MC to include the processes $f\bar{f}\rightarrow f'\bar{f'}, f=\mu,\tau, q,\nu_\ell, f'=e, \mu,\tau, q, \nu_\ell, q=u,d,s,c,b,t, \ell =e,\mu,\tau$ with $f \ne f'$. After giving a brief summary of the CEEX theory with reference to the older EEX (exclusive exponentiation) theory, we illustrate theoretical results relevant to the LHC, FCC-ee, and possible muon collider physics programs.
\end{abstract}

\end{frontmatter}



\section{Introduction}
\label{intro}
In the context of the precision era for QCD in LHC physics ( QCD precision tags $\le 1\%$ ), higher order EW corrections are a necessity, as we have explained in 
Ref.~\cite{kkmc422}. Similarly, the muon collider physics program involves precision studies 
of the properties of the recently discovered~\cite{atlas1,cms1} BEH boson~\cite{beh} and treatment of the effects of higher order EW corrections will be essential, as we illustrate in Ref.~\cite{kkmc422}. Building on our successful YFS/CEEX exponentiation~\cite{yfs,eex,ceex1,ceex2} realization in \KK MC4.13~\cite{kkmc} in precision LEP, B-Factory and Tau-Charm factory physics, we have extended it to \KK MC4.22~\cite{kkmc422} wherein the incoming beams choice, previously restricted  to $e^+,\; e^-$, now allows  $f\bar{f},\; f=e,\mu,\tau, q, \nu_\ell ,\; q=u,d,s,b,\; \ell= e,\mu,\tau$. We note that previous versions of ${\cal KK}$ MC even though not adapted for the LHC were already found useful in estimations of theoretical systematic errors of other calculations~\cite{zwad1,zwad2}.
We also note the approaches of Refs.~\cite{dima,hor,fewz,den-ditt,wac} to EW corrections to heavy gauge boson production at the LHC. In LEP studies~\cite{lepewwg} 
per mille level accuracy required higher order corrections beyond the exact ${\cal O}(\alpha)$ EW corrections . Our studies in ref.~\cite{kkmc422}, briefly exhibited below, show that this is still the case so that the approaches 
in Ref.~\cite{dima,fewz,wac,den-ditt} must be extended to higher orders for precision LHC studies.  Observe that the QED parton shower approach used in Ref.~\cite{hor} for these higher order corrections
is intrinsically a $0-p_T$ formalism and that ad hoc procedures, with varying degrees of success, are used to re-introduce non-zero $p_T$ for the higher order effects whereas our CEEX exponentiation with exact ${\cal O}(\alpha^2L)$ corrections gives the higher effects systematically the non-$p_T$ profile that is exactly correct in the soft limit to all orders in $\alpha$. 
Here, we  give a short summary in the next Section of
the main features of YFS/CEEX exponentiation~\cite{ceex1,ceex2} in the SM EW theory. In Sect.~\ref{reslts} we discuss the changes required to extend the incoming beam choices in the ${\cal KK}$ MC to the more inclusive list of the SM fermions,
present examples of theoretical results
relevant for the LHC, FCC-ee~\cite{fcc-ee} and possible muon collider~\cite{muclldr} 
precision physics programs, and present our summary remarks. 
\section{Review of Standard Model calculations for 
$e^+ e^-$  annihilation with CEEX YFS exponentiation}
\label{review}
We note that  CEEX replaces the older EEX~\cite{eex} -- both are derived from 
the YFS theory~\cite{yfs}. Like what is also now featured in the MC's Herwig++~\cite{hw++} and Sherpa~\cite{shpa} for particle decays, EEX, Exclusive EXponentiation, 
is very close to the original Yennie-Frautschi-Suura formulation. 
CEEX, Coherent EXclusive exponentiation, is actually an extension of the YFS
theory. The coherence of CEEX is friendly to quantum coherence
among the Feynman diagrams: we have the complete
$|\sum_{diagr.}^n {\Meu}_i\big|^2$ rather than the often incomplete
$\sum_{i,j}^{n^2} {\Meu}_i {{\Meu}_j}^*$. The proper treatment of
narrow resonances, $\gamma\oplus Z$~exchanges, $t\oplus s$~channels,
ISR$\oplus$FSR, angular ordering,~etc. are all readily obtained as a consequence. Examples of the EEX formulation are KORALZ/YFS2, BHLUMI, BHWIDE, YFSWW, KoralW and  YFSZZ in our MC event generator approach; the only example of the CEEX formulation is \KK MC.\par
For the process
$e^-(p_1,\lambda_1)+e^+(p_2,\lambda_2) \to f(q_1,\lambda'_1)+\bar{f}(q_2,\lambda'_2)
+\gamma(k_1,\sigma_1)+...+\gamma(k_n,\sigma_n)$ we  illustrate CEEX schematically for 
the full scale CEEX \Order{\alpha^r}, r=1,2, master formula (realized in \KK MC for ISR and FSR up to \Order{\alpha^2}) which
for the polarized total cross section reads 
\begin{align}
\label{eqceex1}
&\sigma^{(r)}\! = \!\!
  \sum_{n=0}^\infty \frac{1}{ n!}
  \int d\tau_{n} ( p_a\!+\!p_b ; p_c,p_d, k_1,\dots,k_n)
\\&~\times
 e^{ 2\alpha\Re B_4 }\!
    \sum_{\sigma_i,\lambda,\bar{\lambda}}\;
    \sum_{i,j,l,m=0}^3
    \hat{\varepsilon}^i_a \hat{\varepsilon}^j_b
    \sigma^i_{\lambda_a \bar{\lambda}_a} 
    \sigma^j_{\lambda_b \bar{\lambda}_b}
\\&~\times
 \Mmf^{(r)}_n 
    \left(\st^{p}_{\lambda} \st^{k_1}_{\sigma_1} \st^{k_2}_{\sigma_2}
                                           \dots \st^{k_n}_{\sigma_n} \right)
    \Big[\Mmf^{(r)}_n\!
     \left(  \st^{p}_{\bar{\lambda}} \st^{k_1}_{\sigma_1} 
             \st^{k_2}_{\sigma_2}\dots \st^{k_n}_{\sigma_n}
     \right)\Big]^\star
 \sigma^l_{\bar{\lambda}_c \lambda_c } 
   \sigma^m_{\bar{\lambda}_d \lambda_d } 
   \hat{h}^l_c              \hat{h}^m_d.
\end{align}

The CEEX amplitudes are

\begin{widetext}{\small
\begin{eqnarray}
\begin{aligned}
\Mmf^{(1)}_n\left(\st^{p}_{\lambda} \st^{k_1}_{\sigma_1}
                 \dots \st^{k_n}_{\sigma_n}\right)\!
  &=\sum\limits_{\wp\in{\cal P}}
   \prod\limits_{i=1}^n  {\sfac}_{[i]}^{\{\wp_i\}}
   \Bigg\{ \hat\beta_0^{(1)}\big(\st^{p}_{\lambda};X_\wp\big)\! 
+\sum\limits_{j=1}^n
   \frac{ \hat\beta^{(1)}_{1\{\wp_j\}}
    \big(\st^{p}_{\lambda}\st^{k_j}_{\sigma_j};X_{\wp}\big)} 
                 { {\sfac}_{[j]}^{\{\wp_j\}} }
   \Bigg\}
\\
\Mmf^{(2)}_n\left(\st^{p}_{\lambda} \st^{k_1}_{\sigma_1}
                              \dots \st^{k_n}_{\sigma_n}\right)
&=\!\sum\limits_{\wp\in{\cal P}}
   \prod\limits_{i=1}^n  {\sfac}_{[i]}^{\{\wp_i\}}
   \Bigg\{ \hat\beta_0^{(2)}\big(\st^{p}_{\lambda};X_\wp\big) 
   \!+\! { \sum\limits_{j=1}^n}
   \frac{\hat\beta^{(2)}_{1\{\wp_j\}}\big(\st^{p}_{\lambda}\st^{k_j}_{\sigma_j};X_{\wp}\big) }
             { {\sfac}_{[j]}^{\{\wp_j\}} }\!
 +\! { \sum\limits_{1\leq j<l\leq n}}
   \frac{\hat\beta^{(2)}_{2\{\wp_j,\wp_l\}}
    \big( \st^{p}_{\lambda}\st^{k_j}_{\sigma_j} \st^{k_l}_{\sigma_l}; X_{\wp}\big)}
          { {\sfac}_{[j]}^{\{\wp_j\}}  {\sfac}_{[l]}^{\{\wp_l\}} }
   \Bigg\}.
\end{aligned}
\end{eqnarray}}
\end{widetext}
See refs.~\cite{ceex1,ceex2} for the detailed IR function definitions  and for many implications such as the \KK MC precision tag $\Delta\sigma/\sigma = 0.2\%$ at LEP established using the results from refs.~\cite{zfitter,berends,recent2}. Extension of \KK MC to the  LHC, FCC and the muon collider avails to their physics such per mille precision.\par
\section{Including the processes {\small
 $f\bar{f}\to f'\bar{f'},$ 
 $f=\mu,q,\nu_\ell$, 
 $f'=\ell, \nu_\ell, q , q=u,d,s,c,b,$
 $\ell=e,\mu,\tau, f\ne f', $} in \KK MC}
\label{reslts}
As we explain in ref.~\cite{kkmc422}, we have  extended the matrix elements, residuals, and IR functions in (\ref{eqceex1}) to the case where we substitute the $e^-,\; e^+$ EW charges by the  EW charges of the new beam particles
$f,\; \bar{f}$ and we substitute the mass $m_e$ everywhere by $m_f$
\footnote{We advise the reader that especially in the QED radiation module KarLud 
for the ISR in \KK MC, see Ref.~\cite{kkmc}, some of the expressions had
$Q_e$ and $m_e$ effectively hard-wired into them and these had to all be found
and substituted properly.}. The resulting new version of \KK MC is version 4.22. 
We have made considerable cross-checks~\cite{kkmc422} both during and after the extension, as we now illustrate.\par
In the most important cross-check, we exhibit in Tab.~\ref{tab:table1} here that, for the $e^+e^-\rightarrow \mu^+\mu^-$ process, 
\KK MC 4.22 reproduces the results in the corresponding $\sqrt{s}=189$GeV
studies done in Ref.~\cite{ceex1} for the dependence of the CEEX calculated cross section and $A_{FB}$ on the energy cut-off on $v=1-s'/s$ where $s'=M^2_{\mu\bar{\mu}}$ is the invariant mass of the $\mu\bar{\mu}$-system. This and its companion results given in ref.~\cite{kkmc422} show that our introduction of the new beams has not spoiled the precision of the \KK MC for the incoming $e^+e^-$ state.
\def\Energy{189GeV}
\def\Process{$e^-e^+ \to \mu^-\mu^+$}
\def\Angle{$\theta^{\bullet}$}
\begin{table}[h]
\centering
{\includegraphics[width=80mm,height=53mm]{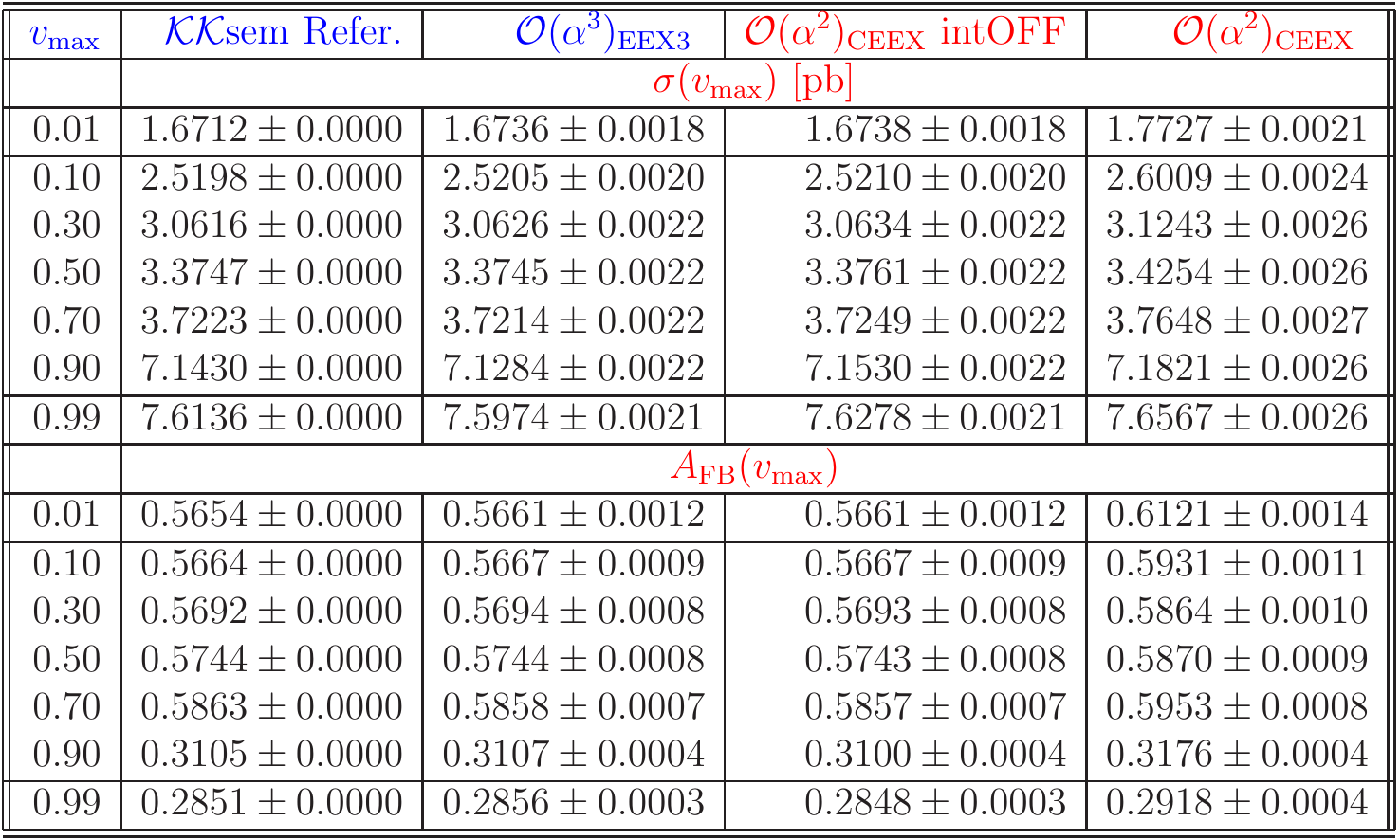}}
\caption{
 Energy cut-off study of
 total cross section  $\sigma$ and charge asymmetry $A_{\rm FB}$
 for annihilation process $e^-e^+ \to \mu^-\mu^+$, at $\sqrt{s}=$189GeV.
 Energy cut: $v<v_{\max}$, $v=1-M^2_{f\bar{f}}/s$.
 Scattering angle for $A_{\rm FB}$ is \Angle
 (defined in Phys. Rev. {\bf D41}, 1425 (1990)).
 No cut in \Angle.
 E-W corr. in \KK\  according to DIZET 6.x~\cite{dima,zfitter}.
 In addition to CEEX matrix element, results are also shown for
 \OrderLL{\alpha^3} EEX3 matrix element without ISR$\otimes$FSR interf.
 \KK{}sem is a semi-analytical program, part of \KK{}MC.
}
\label{tab:table1}
\end{table}
\par
Proceeding pedagogically, especially
given the interest in muon collider precision physics~\cite{muclldr}, we consider next the process $\mu^+\mu^-\rightarrow e^+e^-$ as our first new beam scenario, again 
at $\sqrt{s}=189$GeV to have as a reference the usual
incoming $e^+e^-$ annihilation scenario.  In this new $\mu^+\mu^-$ scenario, while the EW charges are all the same,  the ISR probability to radiate factor $\gamma_e=\frac{2\alpha}{\pi}\left(\ln(s/m_e^2)-1\right)\cong 0.114$ becomes $\gamma_\mu=\frac{2\alpha}{\pi}\left(\ln(s/m_\mu^2)-1\right)\cong 0.0649$. This means that we expect the EW effects where the photonic corrections dominate to show reduction in size for ISR dominated regimes, the same size for the IFI dominated regimes. This is borne-out by the results in Tab. 6 and the companion results
in ref.~\cite{kkmc422}, which together provide a precison tag of 0.2\% at an energy cut of $0.6$. Precision results for EW effects would be available
for the muon collider physics as it will be discussed elsewhere~\cite{elsewh2}. 

\def\Energy{ 189GeV}
\def\Process{$ \mu^- \mu^+ \to e^- e^+$}
%
\par
Turning next to the case of incoming quark anti-quark beams and proceeding as indicated above,  for the process $u \bar{u} \to \mu^-\mu^+$ we obtain the results 
in Tab.~\ref{tab:table3} here and the companion results given in ref.~\cite{kkmc422}, from which we get the precision tag $.08\%$ for the energy cut $0.6$. This satisfies the requirements of precision LHC studies.
\def\Process{$u \bar{u} \to \mu^-\mu^+$}
\begin{table}[h!]
\centering
\includegraphics[width=80mm,height=53mm]{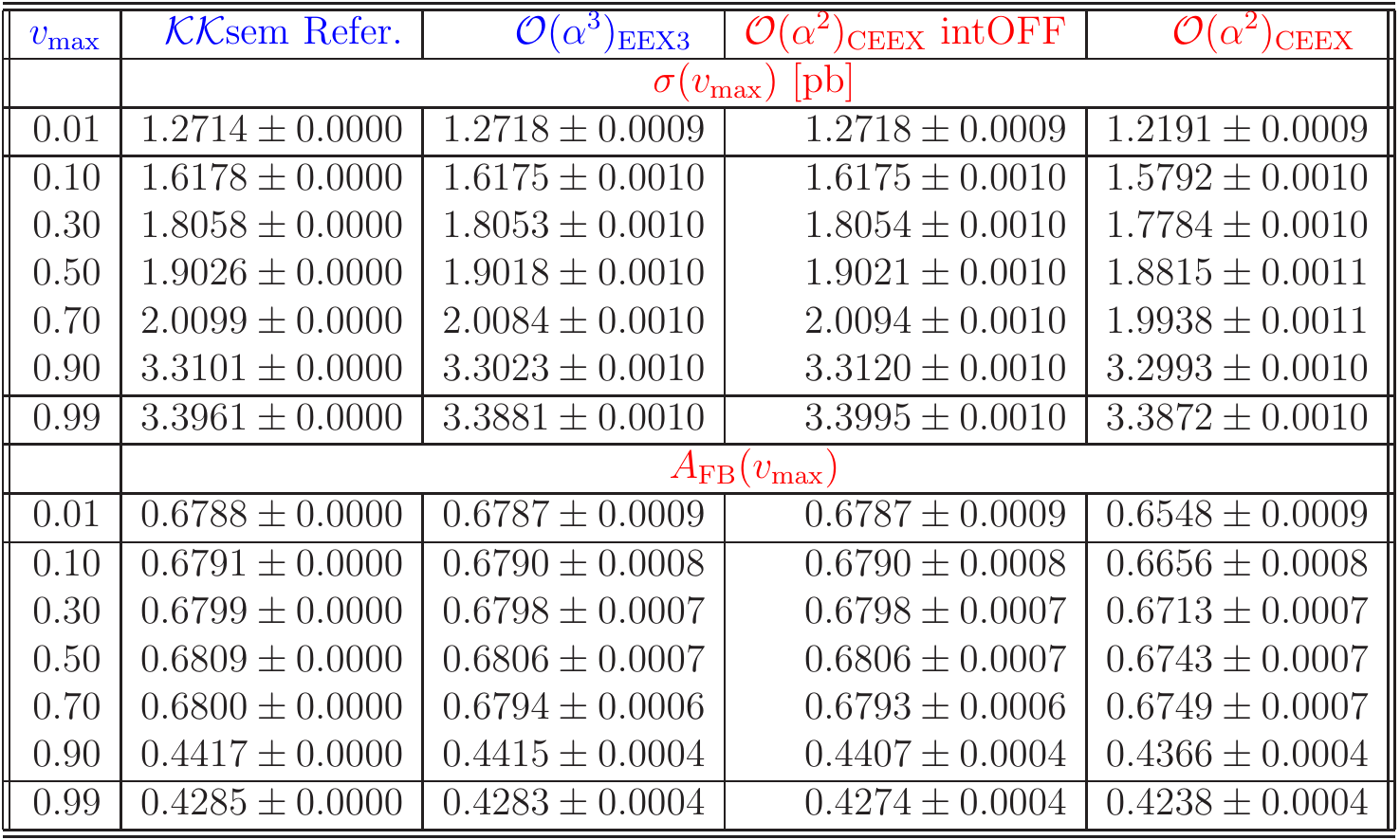}
\caption{
 Study of total cross section $\sigma(v_{\max})$ 
 and charge asymmetry $A_{\rm FB}(v_{\max})$,
 \Process, at $\sqrt{s}$~=\Energy.
 See Table \ref{tab:table1} for definition of
 the energy cut $v_{\max}$, scattering angle and M.E. type, 
}
\label{tab:table3}
\end{table}
\par
The third step in our extension is the introduction of PDF's for the quark beams. This is currently done in a hard-wired way using the beamsstrhalung module in the \KK MC. A more
algorithmic formulation of this part of the extension is in progress~\cite{elsewh2}. In Appendix A of ref.~\cite{kkmc422}, 
sample output (three events in the LUND MC format) is given from a run of \KK\ MC version 4.22 
for $pp\to u\bar{u} \to l^- l^+ +n\gamma$ where
simple parton distribution functions (PDF's)
of $u$ and $\bar{u}$ quarks
in the proton are replacing beamsstrahlung distributions
(see function {\tt BornV\_RhoFoamC} in the source code). 
The proton remnants are now represented by two photons in the event record with the 
exactly zero transverse momentum which were
formerly beamsstrahlung photons (temporary fix).  
The multiple photon event in the sample output with a photon of  $p_T = 31.6$ GeV emphasizes the need to use \KK MC 4.22 to address the systematics of the current treatment of QED ISR at LHC with QED PDF's~\cite{elsewh2}. See ref.~\cite{kkmc422} for further discussion of this sample output.
\par
\begin{figure}[h]
\centering
\includegraphics[width=70mm]{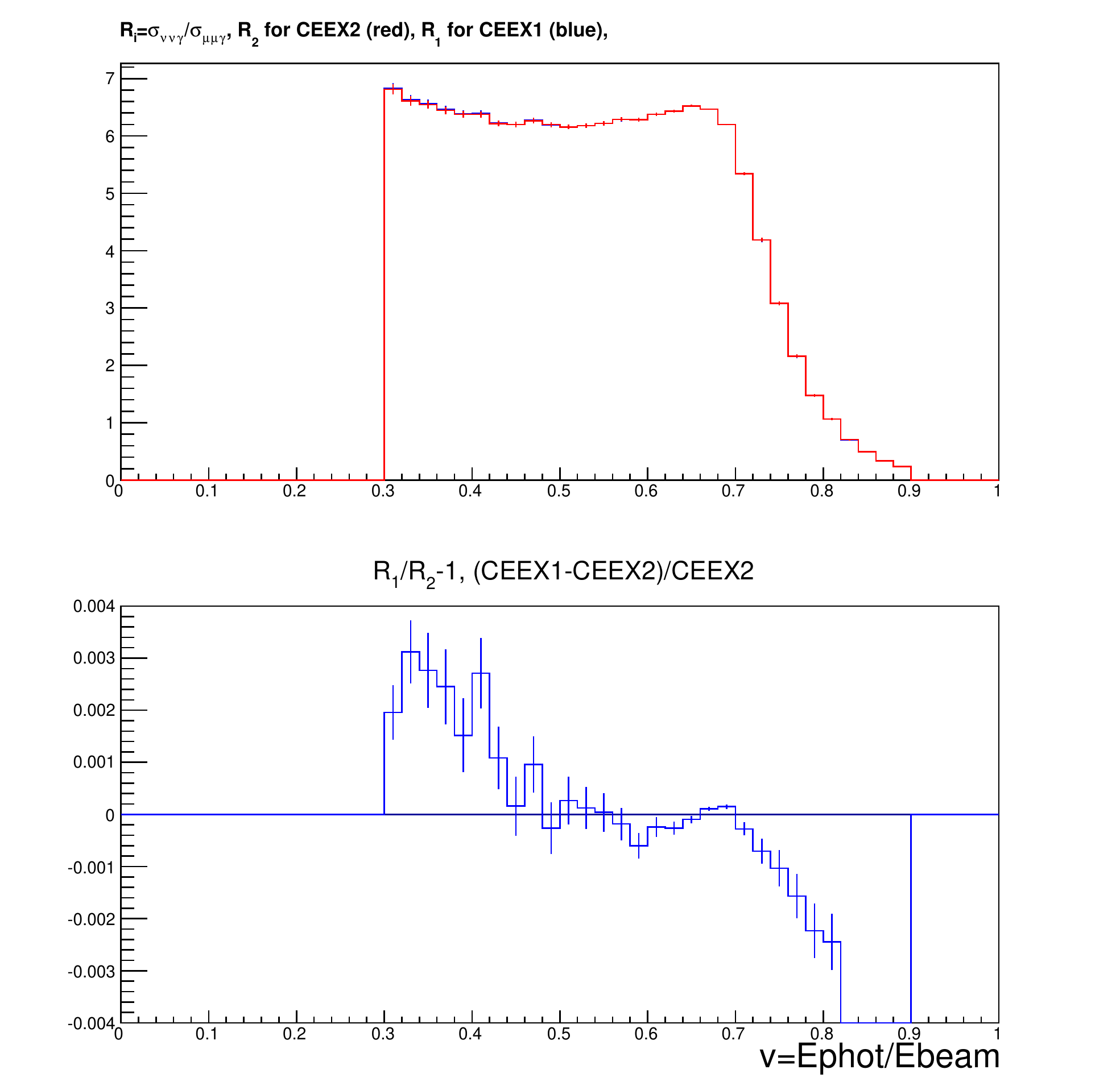}
\caption{ QED correction estimate in $\sigma(\nu\bar\nu\gamma)/\sigma(\mu\bar\mu\gamma)$: QED corrections seem to cancel dramatically in this ratio and drop to $\sim 0.03\%$. 
This preliminary result is undergoing further tests~\cite{elsewh2}.
}
\label{figNnu}
\end{figure}

Further development of \KK MC, such as its inclusion in QCD parton shower MC's~\cite{herwiri2},  is in progress as discussed in ref.~\cite{kkmc422}. Here we would also note application of \KK MC as it stands to 
new colliding beam devices such as FCC-ee~\cite{fcc-ee}, where we show in Fig.~\ref{figNnu} the QED correction predicted by \KK MC for $\sigma(\nu\bar\nu\gamma)/\sigma(\mu\bar\mu\gamma)$ as a probe of the invisible Z width at 161 GeV with the cuts as prescribed in ref.~\cite{tlep-study}. 
The QED corrections seem to cancel to $\sim 0.03\%$: in the Z return regime, we get the initial theoretical precision error estimate $\Delta N^{th}_\nu \cong 0.00045$, consistent with the goals of the FCC-ee~\cite{tlep-study}. Further studies are needed to finalize this result~\cite{elsewh2}. To sum up, \KK MC is still alive and still useful. We thank Prof. I. Antoniadis for the support of the CERN TH Unit while this work was completed.This work is partly supported by
the Polish National Science Centre grant DEC-2011/03/B/ST2/02632. \par












\end{document}